\documentclass{article}
\usepackage{float}
\usepackage[letterpaper]{geometry}

\pdfoutput=1
\usepackage{hyperref}
\hypersetup{
  pdfinfo={
    Title={Your title hear},
    Author={Castaly Fan},
    Subject={If you want to put something here, do so},
    Keywords={Add some keywords if you feel so inclined}
  }
}
\usepackage{pdfpages}

\usepackage[utf8]{inputenc}
\usepackage{graphicx}
\usepackage[justification=centering]{caption}
\usepackage{subcaption}

\usepackage{listings}
\usepackage{xcolor}
\definecolor{codegreen}{rgb}{0,0.6,0}
\definecolor{codegray}{rgb}{0.5,0.5,0.5}
\definecolor{codepurple}{rgb}{0.58,0,0.82}
\definecolor{backcolour}{rgb}{0.95,0.95,0.92}

\lstdefinestyle{mystyle}{
    backgroundcolor=\color{backcolour},   
    commentstyle=\color{codegreen},
    keywordstyle=\color{magenta},
    numberstyle=\tiny\color{codegray},
    stringstyle=\color{codepurple},
    basicstyle=\ttfamily\footnotesize,
    breakatwhitespace=false,         
    breaklines=true,                 
    captionpos=b,                    
    keepspaces=true,                 
    numbers=left,                    
    numbersep=5pt,                  
    showspaces=false,                
    showstringspaces=false,
    showtabs=false,                  
    tabsize=2
}

\title{A Brief Note of Analyzing and Plotting $\nu_\mu$ Disappearance in SBN Detector under ROOT Framework}
\author{Castaly Fan}
\date{January 2020}

\begin{document}
\maketitle

\begin{abstract}
    This is a brief technical note of analyzing the $\nu_\mu$ disappearance in SBN detector. We here provide a kind of method of plotting the histograms and the heat map plot. We will explore the properties via ROOT framework supported by CERN. The main language we used is C++.
\end{abstract}

\tableofcontents
\clearpage

\section{Background}
For searching sterile neutrinos, one way is to look over the disappearence (reduction) of muon neutrinos ($\nu_\mu$) in a muon neutrino beam. The data of detector properties we have is shown as below:
\begin{table}[H]
  \begin{center}
    \caption{Data of detector properties\cite{1}}
    \label{tab:table1}
    \begin{tabular}{c|l|l|l} 
      Detector & Baseline & Active LAr mass & POT (protons on target) \\
      \hline
      SBND & 110m & 112 tons & $6.2 \times10^{20}$\\
      MicroBooNE & 470m & 89 tons & $13.2 \times10^{20}$\\
      ICARUS & 600m & 476 tons & $6.2 \times10^{20}$\\
    \end{tabular}
  \end{center}
\end{table}

The cross section $\sigma (E_{\nu})$ for CCQE $ \nu_{mu}$ interactions on $^{4}_{}Ar$ is shown as the left figure below. And the initial neutrino flux $\phi (E_{\nu})$ without oscillations at MicroBooNE is shown as the right figure below.
\begin{figure}[H]
  \centering
  \begin{subfigure}[b]{0.45\linewidth}
    \includegraphics[width=1.0\linewidth]{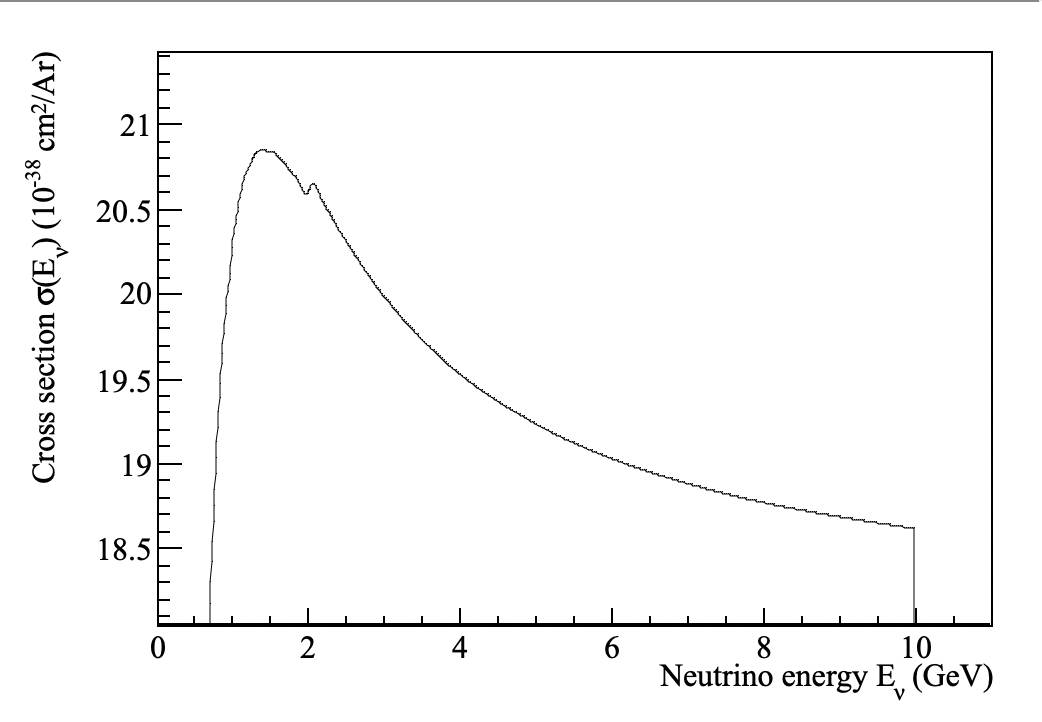}
    \caption{cross section}
  \end{subfigure}
  \begin{subfigure}[b]{0.45\linewidth}
    \includegraphics[width=1.0\linewidth]{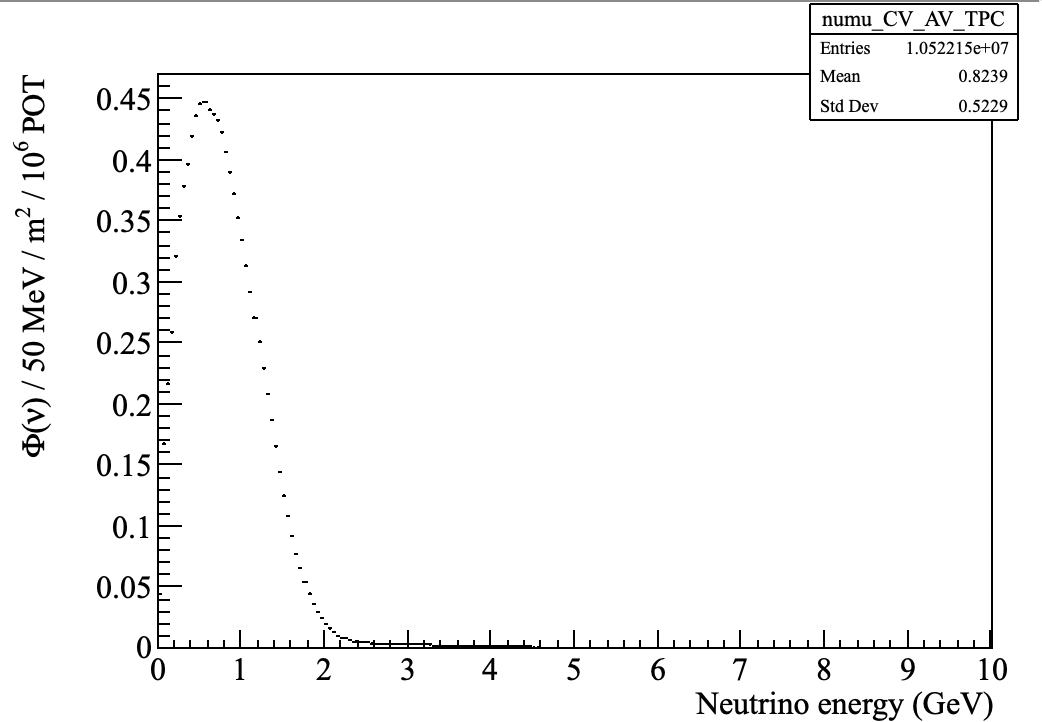}
    \caption{flux at MicroBooNE}
  \end{subfigure}
  \caption{The figure of cross section and flux.}
  \label{fig 1}
\end{figure}

Given the number of argon nuclei $N_{Ar}$, we can calculate the event rate
    \begin{equation}
	N_{ev}(E_{\nu}) = \sigma(E_{\nu}) \times N_{Ar} \times [\phi(E_{\nu}) \times \Delta E_{\nu} ]\times POT
	\label{eq 1}
    \end{equation}

Armed with the equation \ref{eq 1} and the data of Table \ref{tab:table1}, we can start the task step by step.

\section{Histograms of the Event Rate}
Equation \ref{eq 1} shows that the event rate depends on the number of $4_{Ar}$, which can be expressed as
    \begin{equation}
    N_{Ar} = \frac{M \times N_{A} }{W} 
    \label{eq 2}
    \end{equation}
where $M$ is the mass in gram, $W$ is the atomic weight, and $N_{A}$ is the number of nuclei per mole (namely, $6.02\times10^{23}$).
Now, let us consider $N_{Ar}$ of each detectors. Given the LAr mass of Table \ref{tab:table1}, we obtain:
\begin{itemize}
    \item $1.6856\times10^{30}$ for SBND
    \item $1.3395\times10^{30}$ for MicroBooNE
    \item $7.1638\times10^{30}$ for ICARUS
\end{itemize}
Then, we can create the event rate's histogram according to Equation \ref{eq 1}.

Back to the Figure \ref{fig 1}, the cross section for each detector is the same, whereas the curve of the flux would be scaled by the factor $\frac{1}{r^{2}}$ since the flux falls off with the distance r from the target. For instance, given that the flux at MicroBooNE depends on the distance 470 m, the flux at SBND would be scaled by $(470/110)^2$ m and the flux at ICARUS would be scaled by $(470/600)^2$ m.

Given that Equation \ref{eq 1}, the unit should be considered in order to create the event rate's histogram. So we can simply write down the event rate equation in terms of the units:
    \begin{equation}
    [N] = [\frac{cm^2}{Ar}]\times[Ar]\times[\frac{1}{MeV\cdot POT \cdot m^2}]\times[MeV]\times[POT]
    \label{eq 3}
    \end{equation}

\begin{figure}[H]
    \centering
    \includegraphics[width=0.85\textwidth]{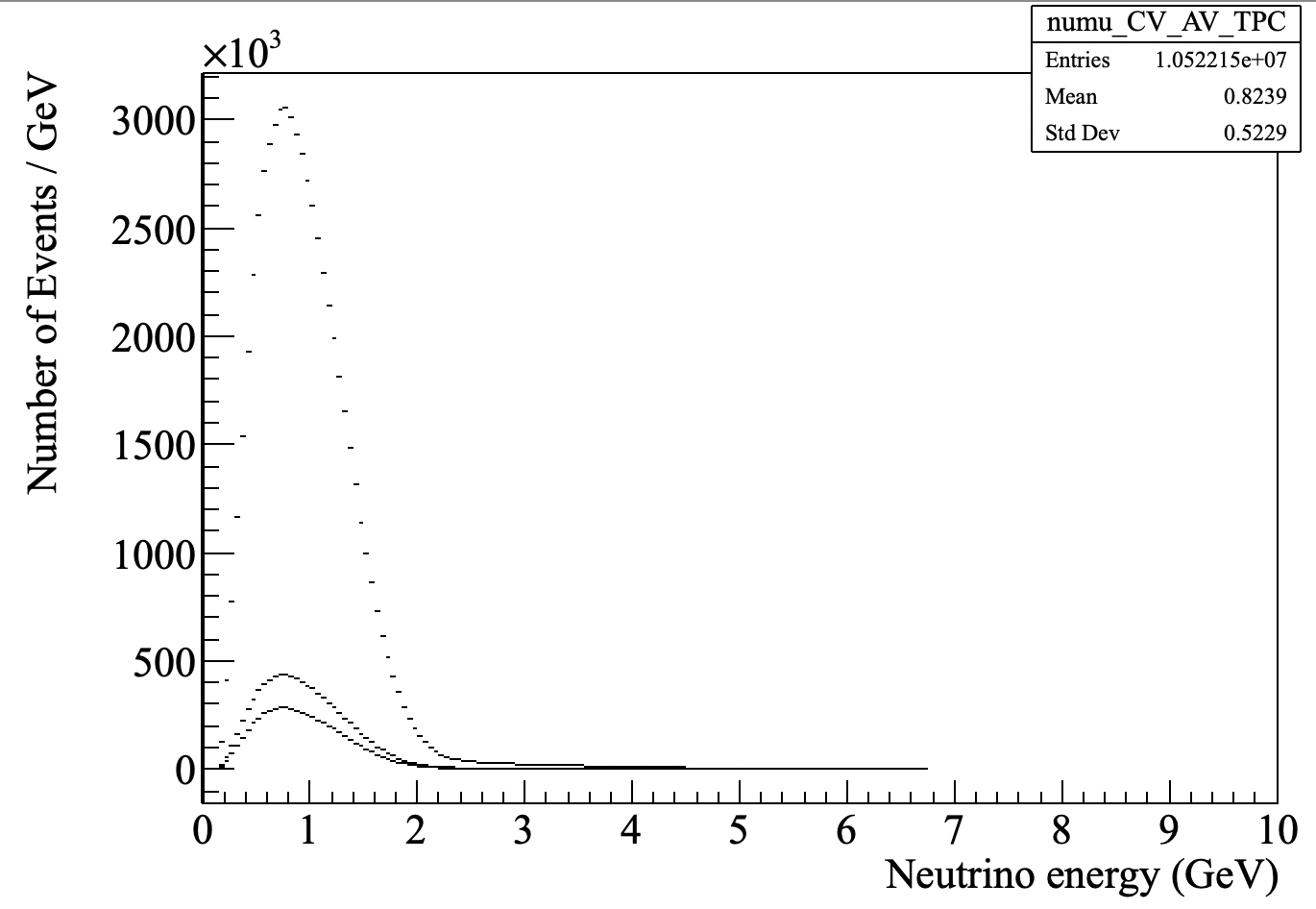}
    \caption{Event rate's histogram. The curves from up to down are respectively corresponding to SBND, ICARUS, and MicroBooNE.}
    \label{fig 2}
\end{figure}
Here, $[N]$ is denoted by the number of event. Then we have to notice the unit should be normalized, in particular, $[cm^2]$ is for cross section whereas $[m^2]$ is for flux. 

As a result (See Appendix 1), the histogram of event rate for each detector can be created as Figure \ref{fig 2}.

By calculating the total number of events according to the Figure\ref{fig 2}, we can obtain the value $3.13398\times10^6$, which is almost matched with the number of $\nu_\mu$ events during $\nu_{\mu} n \rightarrow \mu^{-}p$ process based on the SBN proposal (namely, 3,122,600).

\section{Neutrino Oscillations}
To explore the disappearance of $\nu_\mu$, we have to know the probability where $\nu_\mu$ remain, denoted by $P_{\nu_\mu \rightarrow \nu_\mu}$. In general, if $P_{\nu_\mu \rightarrow \nu_x}$ means the probability that $\nu_\mu$ turn into $\nu_e$, $\nu_\tau$, or $\nu_s$ (sterile neutrinos), then $P_{\nu_\mu \rightarrow \nu_\mu}$ would be reasonably equal to $1-P_{\nu_\mu \rightarrow \nu_x}$. Specifically, we have a formula describing $\nu_\mu$ disappearance as below\cite{2}:
    \begin{equation}
        P^{3+1}_{\nu_\mu \rightarrow \nu_\mu} = 1-\sin^2(2\theta_{\mu\mu})\sin^2\left(\frac{\Delta m^2 L}{4E_\nu}\right)
        \label{eq 4}
    \end{equation}
The equation above is taken by natural units. Thus, we can write down the form in SI units:

\begin{equation}
        P^{3+1}_{\nu_\mu \rightarrow \nu_\mu} = 1-\sin^2(2\theta_{\mu\mu})\sin^2\left(1.27\frac{\Delta m^2 L}{E_\nu}\right)
        \label{eq 5}
    \end{equation}

where $\theta$ is the mixing angle, $L$ is the oscillation distance in kilometers, and $E$ is the neutrino energy in GeV.
\begin{figure}[H]
    \centering
    \includegraphics[width=0.85\textwidth]{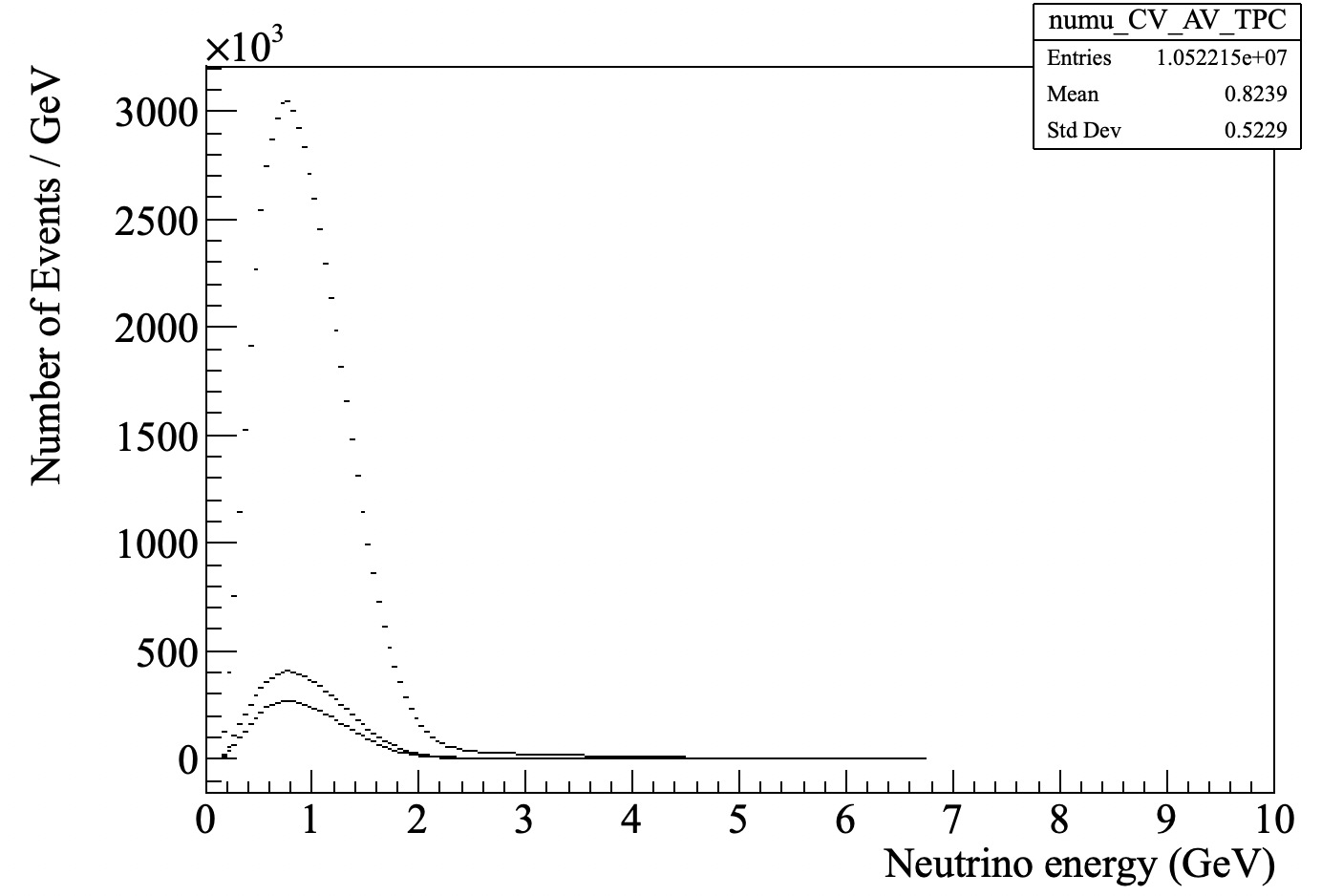}
    \caption{Event rate's histogram with oscillations. The curves of SBND, ICARUS, and MicroBooNE are same as Fig \ref{fig 2}, whereas have subtle differences between both diagrams.}
    \label{fig 3}
\end{figure}
By arbitrarily setting $\sin^2(2\theta_{\mu \mu})=0.1$ and $\Delta m^2=1 eV$, the factor of probability can be simply written as
\begin{equation}
    P^{3+1}_{\nu_\mu \rightarrow \nu_\mu}=1-0.1\sin^2\left(1.27\frac{ L}{E_\nu}\right)
    \label{eq 6}
\end{equation}
, which shows the probability of oscillations depends on the distance $L$ and the neutrinos energy $E_{\nu}$.

To create the event rate histogram with neutrino oscillations, we can consider the factor of equation \ref{eq 6}. The distance $L$ is corresponding to each detector's baseline (see Table \ref{tab:table1}), as for the energy $E_{\nu}$, it is given by the flux histogram as Figure \ref{fig 1}. Now, we can obtain the event rate histogram with oscillations, which is shown as Figure \ref{fig 3} (See Appendix 2).

By looking at the Figure \ref{fig 3} roughly, it is hard to see obvious differences between that with the one without oscillations (Figure \ref{fig 2}). However, there are actually few subtle differences between both histograms. Now what we have to do is to analyze the details of both curves via statistical ways.

\section{Statistical Testing}
Statistically, to do data fitting work, we need to use least squares in order to minimize the sum of squared errors and residuals\cite{3}. Consider a function $y_{i}, i=1,...,N$ as a function of another variable $x_{i}$ without error. Each $y_{i}$ has a different unknown mean $\lambda_{i}$ and a known variance $\sigma^{2}_{i}$. Then, suppose the true value $\lambda$ is the function of $x$ and depends on unknown parameters $\theta=(\theta_{1}, ..., \theta_{m})$, that is, $\lambda=\lambda(x;\theta)$. Now, consider the $\chi^2(\theta)$ quantity:
    \begin{equation}
        \chi^2(\theta)=\sum_{i=1}^{N} \frac{(y_{i}-\lambda(x_{i};\theta))^2}{\sigma^2_i}
    \label{eq 7}
    \end{equation}
In other words, $y_{1},...,y_{N}$ are measured with errors $\sigma_{1},...,\sigma_{N}$ at each $x (x_{1},...,x_{N})$ value without errors. The true value $\lambda_{i}=\lambda(x_{i};\theta)$, where $\theta$ is adjusted to minimize the value of $\chi^2$ given by the equation above. 

Assume un-oscillated event rate histogram is $y_i$ and oscillated one is another function $\lambda$. For the variance $\sigma^{2}_{i}$, consider it as $\lambda$, since we regard the case as a Poisson distribution (i.e. the mean equal to the variance). Thus, the chi-squared value of one detector can be simply obtained from:
    \begin{equation}
        \chi^2=\sum_{i=1}^{N} \frac{(y_{i}-\lambda_{i})^2}{\lambda_i}
    \label{eq 8}
    \end{equation}
As for the total chi-squared value for three detectors can be written as the sum
    \begin{equation}
        \chi^2_{total}=\chi^2_{a}+\chi^2_{b}+\chi^2_{c}
    \label{eq 9}
    \end{equation}
where the index a, b, and c can be represented as the detector SBND, MicroBooNE, and ICARUS respectively. 

Then we can print out the $\chi^2$ value for each detector (Appendix 3). we get:

    \begin{itemize}
        \item SBND: $\chi^2_a=0.278707$
        \item MicorBooNE: $\chi^2_b=1.54599$
        \item ICARUS: $\chi^2_c=3.45765$
    \end{itemize}

So that the total $\chi^2$ value is $\chi^2_{total}=\chi^2_{a}+\chi^2_{b}+\chi^2_{c}=5.282347$, which shows the difference between null hypothesis (un-oscillated) with oscillated histograms.

\section{The Heat Map Plot}
Given that the $\chi^2$ value for each detector, we can make a heat map plot to show whether the experiment can be consistent with or rule out statistically. To realize it, we must make a 2-D plot, that is, TH2D in ROOT framework (Appendix 4). 

\begin{figure}[H]
    \centering
    \includegraphics[width=0.90\textwidth]{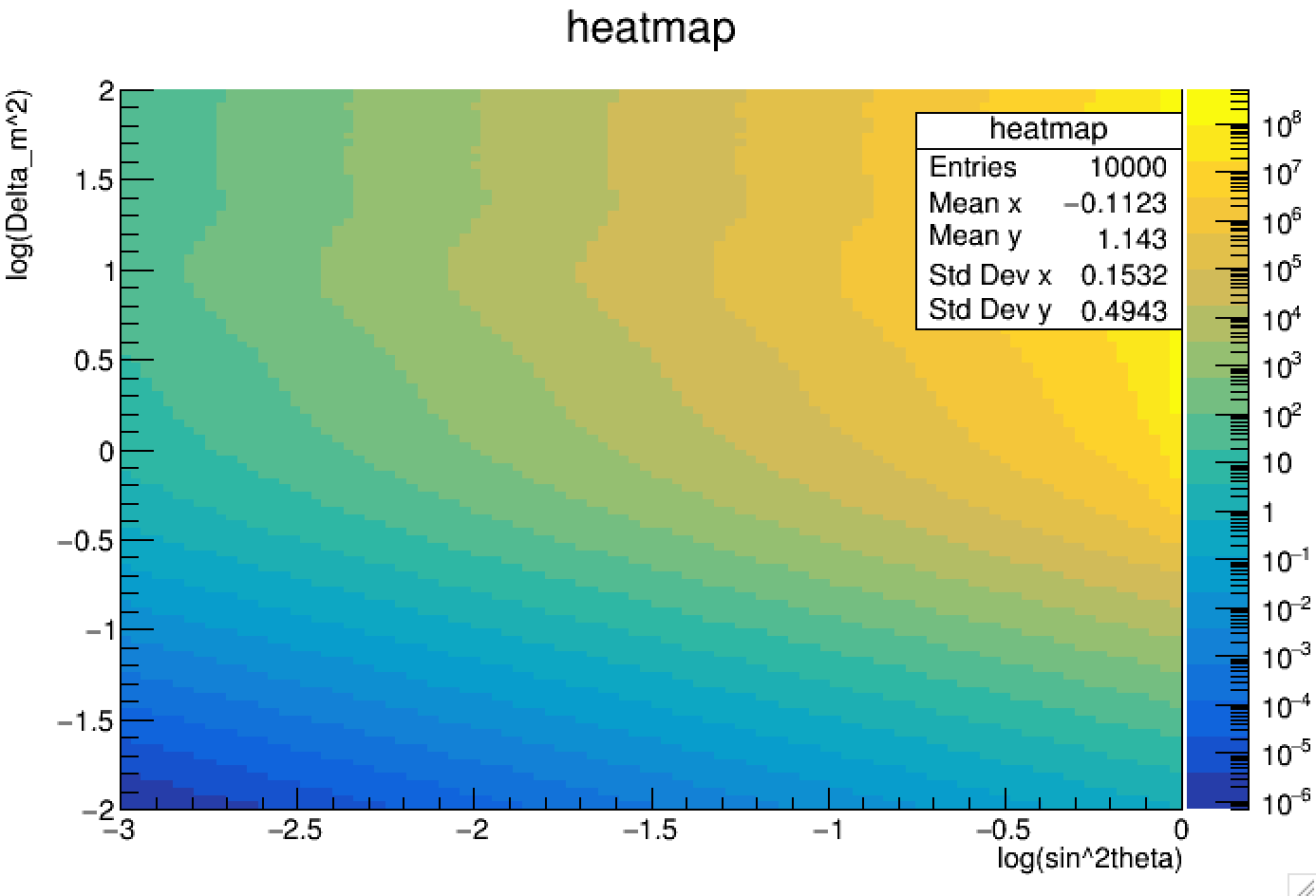}
    \caption{Heat map plot given from each detector's $\chi^2$ value. The contour curves can be seen as the sensitivity prediction for the detectors.}
    \label{fig 4}
\end{figure}

\begin{figure}[H]
    \centering
    \includegraphics[width=0.75\textwidth]{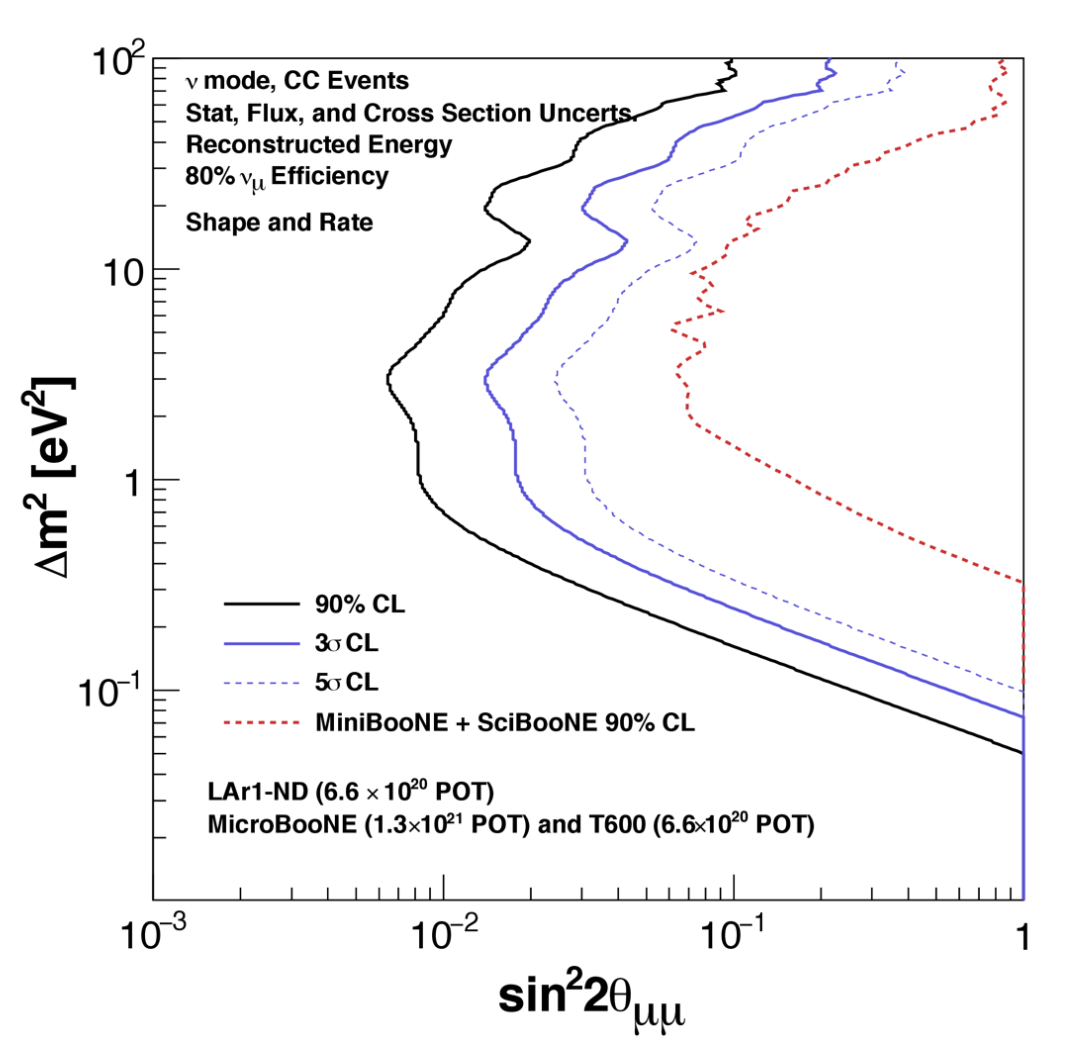}
    \caption{The formal heat map plot in SBN proposal (arxiv:1903.04608). Compared with both plots, the curves are almost similar so that the result would be correct\cite{2}.}
    \label{fig 5}
\end{figure}

Ultimately, the plot on canvas is shown as Figure \ref{fig 4}. The amazing result is important in testing the sensitivity for each detector. Note that the axis is set as logarithmic value. For the formal heat map plot, the curves can be shown more obviously, which is shown as Figure \ref{fig 5}.

\section{Conclusion}
To create the 2-D heat map plot, the first step is to calculate event rate according to cross section's and flux's data. Then, considering neutrino oscillations, we can see the slightly different between both event rate's histograms. 
To know the difference between null hypothesis (non-oscillated) with the  oscillated one, we must figure out the $\chi^2$ value. Finally, compiling the result into the original loop, we can attain the heat map plot with the curves.
It is noteworthy that physicists can look over the sensitivity from such contour plot. In this way, it can let us know the experimental details of $\nu_{\mu}$ disappearance in SBN detector.

\section{Acknowledgement}
I am grateful to Prof. Andrew Mastbaum for giving me a precious opportunity to be engaged in SBN program's research. It is a wonderful tour for me to learn plenty of technological skills as well as data analysis work, which is crucial to my physics career in the near future.

\clearpage

\section{Appendices}
Here is the collection of some related codes with respect to ROOT framework.
\subsection{Appendix 1: Event rate histograms}
To create the plot for event rate histograms, we can log into the ROOT framework and run the following command (some "//" means the note for the command line below):
\begin{lstlisting}[language=C++]
void draw_histogram(TH1D* h, TGraph* cross_section, double parameters)
{
	std::cout << parameters << std::endl;
	// for every bin in x-axis (GeV)
	for (int i = 1; i <= h->GetNbinsX(); ++i)
	{
		float binCenter = h->GetBinCenter(i);

		// value of cross section 
		float y1 = cross_section->Eval(binCenter);

		// value of flux 
		float y2 = h->GetBinContent(i);

		// eventrate formula
		float N = y2 * y1 * parameters;

		// assign the y value back to the plot
		h->SetBinContent(i,N);
		
		// debug output
		// std::cout << N << std::endl;
	}
	h->GetYaxis()->SetTitle("Number of Events / GeV");
	h->Draw("SAME");
	std::cout << h->Integral("width") << std::endl;
}

void eventrate()
{
	// find the directory where cross section file is 
	TFile* cross_section_file = TFile::Open("cross_section.root");
	cross_section_file->ls();

	// read the cross section file 
	TGraph* cross_section = (TGraph*) cross_section_file->Get("qel_cc_n;1");
	
	// draw the cross section figure 
	// cross_section->Draw();

	// find the directory where flux file is 
	TFile* flux = TFile::Open("flux.root");
	flux->ls();
	
	// read the flux file
	TH1D* h = (TH1D*) flux->Get("numu_CV_AV_TPC");

	// create three clones for three cases
	TH1D* h_sbnd = (TH1D*) h->Clone("h_sbnd");
	TH1D* h_microboone = (TH1D*) h->Clone("h_microboone");
	TH1D* h_icarus = (TH1D*) h->Clone("h_icarus");
	
	draw_histogram(h_sbnd, cross_section,
	            1E-38/1E4*1.6856E30/0.05*6.2E20/1E6*(470.0/110)*(470.0/110));
	draw_histogram(h_microboone, cross_section,
	            1E-38/1E4*1.3395E30/0.05*13.2E20/1E6);
	draw_histogram(h_icarus, cross_section,
	            1E-38/1E4*7.1638E30/0.05*6.2E20/1E6*(470.0/600)*(470.0/600));
}
\end{lstlisting}

\subsection{Appendix 2: Event rate histograms with oscillations}
The command here is slightly different from the first one, that is, we have introduced "baseline" as a new parameter:
\begin{lstlisting}[language=C++]
void draw_histogram(TH1D* h, TGraph* cross_section,
                    double parameters, double baseline)
{
	std::cout << parameters << std::endl;
	for (int i = 1; i <= h->GetNbinsX(); ++i)
	{
		double binCenter = h->GetBinCenter(i);
		double y1 = cross_section->Eval(binCenter);
		double y2 = h->GetBinContent(i);
	
		// eventrate formula
		double S = sin(1.27/1E3 * baseline/binCenter);
		//std::cout << i << ": " << y2 << std::endl;
		double N = (1.0 - (0.1) * S * S )* y1 * y2 * parameters;

		h->SetBinContent(i,N);
		// std::cout << N << std::endl;
	}
	h->GetYaxis()->SetTitle("Number of Events / GeV");
	h->Draw("SAME");
	std::cout << h->Integral("width") << std::endl;
}

void eventrate2()
{
	TFile* cross_section_file = TFile::Open("cross_section.root");
	cross_section_file->ls();
	TGraph* cross_section = (TGraph*) cross_section_file->Get("qel_cc_n;1");
	// cross_section->Draw();

	TFile* flux = TFile::Open("flux.root");
	flux->ls();
	TH1D* h = (TH1D*) flux->Get("numu_CV_AV_TPC");

	TH1D* h_sbnd = (TH1D*) h->Clone("h_sbnd");
	TH1D* h_microboone = (TH1D*) h->Clone("h_microboone");
	TH1D* h_icarus = (TH1D*) h->Clone("h_icarus");
	
	draw_histogram(h_sbnd, cross_section, 
		1E-38/1E4*1.6856E30/0.05*6.2E20/1E6*(470.0/110.0)*(470.0/110.0),
		110.0);
	draw_histogram(h_microboone, cross_section, 
		1E-38/1E4*1.3395E30/0.05*13.2E20/1E6,
		470.0);
	draw_histogram(h_icarus, cross_section, 
		1E-38/1E4*7.1638E30/0.05*6.2E20/1E6*(470.0/600)*(470.0/600),
		600.0);
}
\end{lstlisting}

\subsection{Appendix 3: $\chi^{2}$ values}
Here, we need to introduce the parameter s2 ($sin^2 \theta$) and m (mass):
\begin{lstlisting}[language=C++]
void draw_histogram(TH1D* h, TGraph* cross_section, double parameters, 
                    double baseline, double s2, double m)
{
	std::cout << parameters << std::endl;
	for (int i = 1; i <= h->GetNbinsX(); ++i)
	{
		double binCenter = h->GetBinCenter(i);
		double y1 = cross_section->Eval(binCenter);
		double y2 = h->GetBinContent(i);
		double S = sin(1.27/1E3 * m * baseline/binCenter);
		//std::cout << i << ": " << y2 << std::endl;
		double y = y1 * y2 * parameters;
		double N = (1.0 - s2 * S * S )* y;
		h->SetBinContent(i,N);
		// std::cout << N << std::endl;
	}
	h->GetYaxis()->SetTitle("Number of Events / GeV");
	h->Draw("SAME");
	std::cout << h->Integral("width") << std::endl;
}

double chi2(TH1D* h, TH1D* g)
{
	double chi_sq = 0.0;
	//for (int i = 1; i <= h->GetNbinsX(); ++i)
	for (int i = 1; i <= 135; ++i)
	{
		double y = h->GetBinContent(i);
		double lambda = g->GetBinContent(i);
		if (lambda == 0) continue;
		double chi_sq_i = (y-lambda)*(y-lambda)/lambda;
		std::cout << "#" << i << "y:" << y << ", lambda:" 
		           << lambda << std::endl;
		
		chi_sq = chi_sq + chi_sq_i;
	}
	return chi_sq;
}

void eventrate_chi()
{
	TFile* cross_section_file = TFile::Open("cross_section.root");
	cross_section_file->ls();

	TGraph* cross_section = (TGraph*) cross_section_file->Get("qel_cc_n;1");
	// cross_section->Draw();

	TFile* flux = TFile::Open("flux.root");
	flux->ls();
	TH1D* h = (TH1D*) flux->Get("numu_CV_AV_TPC");

	TH1D* h_sbnd = (TH1D*) h->Clone("h_sbnd");
	TH1D* h_microboone = (TH1D*) h->Clone("h_microboone");
	TH1D* h_icarus = (TH1D*) h->Clone("h_icarus");
	
	draw_histogram(h_sbnd, cross_section, 
		1E-38/1E4*1.6856E30/0.05*6.2E20/1E6*(470.0/110.0)*(470.0/110.0),
		110.0, 0, 1);
	draw_histogram(h_microboone, cross_section, 
		1E-38/1E4*1.3395E30/0.05*13.2E20/1E6,
		470.0, 0, 1);
	draw_histogram(h_icarus, cross_section, 
		1E-38/1E4*7.1638E30/0.05*6.2E20/1E6*(470.0/600)*(470.0/600),
		600.0, 0, 1);

	TH1D* h_sbnd2 = (TH1D*) h->Clone("h_sbnd2");
	TH1D* h_microboone2 = (TH1D*) h->Clone("h_microboone2");
	TH1D* h_icarus2 = (TH1D*) h->Clone("h_icarus2");
	
	draw_histogram(h_sbnd2, cross_section, 
		1E-38/1E4*1.6856E30/0.05*6.2E20/1E6*(470.0/110.0)*(470.0/110.0),
		110.0, 0.001, 1);
	draw_histogram(h_microboone2, cross_section, 
		1E-38/1E4*1.3395E30/0.05*13.2E20/1E6,
		470.0, 0.001, 1);
	draw_histogram(h_icarus2, cross_section, 
		1E-38/1E4*7.1638E30/0.05*6.2E20/1E6*(470.0/600)*(470.0/600),
		600.0, 0.001, 1);

	double chi2_sbnd = chi2(h_sbnd, h_sbnd2);
	double chi2_microboone = chi2(h_microboone, h_microboone2);
	double chi2_icarus = chi2(h_icarus, h_icarus2);

	std::cout << "chi2_sbnd: " << chi2_sbnd << std::endl;
	std::cout << "chi2_microboone: " << chi2_microboone << std::endl;
	std::cout << "chi2_icarus: " << chi2_icarus << std::endl;

}
\end{lstlisting}

\subsection{Appendix 4: Heat map plot}
Armed with the event rate histograms as well as the $\chi^2$ values, we can write down the loop in our original function. The entire command lines are shown as below:

\begin{lstlisting}[language=C++]
void draw_histogram(TH1D* h, TGraph* cross_section, double parameters, 
                    double baseline, double s2, double m)
{
	//std::cout << parameters << std::endl;
	for (int i = 1; i <= h->GetNbinsX(); ++i)
	{
		double binCenter = h->GetBinCenter(i);
		double y1 = cross_section->Eval(binCenter);
		double y2 = h->GetBinContent(i);

		double S = sin(1.27/1E3 * m * baseline/binCenter);
		//std::cout << i << ": " << y2 << std::endl;
		double y = y1 * y2 * parameters;
		double N = (1.0 - s2 * S * S )* y;
		h->SetBinContent(i,N);
		// std::cout << N << std::endl;
	}
	h->GetYaxis()->SetTitle("Number of Events / GeV");
	h->Draw("SAME");
	//std::cout << h->Integral("width") << std::endl;
}

double chi2(TH1D* h, TH1D* g)
{
	double chi_sq = 0.0;
	//for (int i = 1; i <= h->GetNbinsX(); ++i)
	for (int i = 1; i <= 135; ++i)
	{
		double y = h->GetBinContent(i);
		double lambda = g->GetBinContent(i);
		if (lambda == 0) continue;
		double chi_sq_i = (y-lambda)*(y-lambda)/lambda;
		std::cout << "#" << i << " y: " << y << ", lambda: "
		           << lambda << std::endl;
		chi_sq = chi_sq + chi_sq_i;
	}
	return chi_sq;
}

void eventrate_heatmap()
{
	TFile* cross_section_file = TFile::Open("cross_section.root");
	cross_section_file->ls();
	TGraph* cross_section = (TGraph*) cross_section_file->Get("qel_cc_n;1");
	// cross_section->Draw();
	TFile* flux = TFile::Open("flux.root");
	flux->ls();
	TH1D* h = (TH1D*) flux->Get("numu_CV_AV_TPC");

	TH1D* h_sbnd = (TH1D*) h->Clone("h_sbnd");
	TH1D* h_microboone = (TH1D*) h->Clone("h_microboone");
	TH1D* h_icarus = (TH1D*) h->Clone("h_icarus");
	
	draw_histogram(h_sbnd, cross_section, 
		1E-38/1E4*1.6856E30/0.05*6.2E20/1E6*(470.0/110.0)*(470.0/110.0),
		110.0, 0, 1);
	draw_histogram(h_microboone, cross_section, 
		1E-38/1E4*1.3395E30/0.05*13.2E20/1E6,
		470.0, 0, 1);
	draw_histogram(h_icarus, cross_section, 
		1E-38/1E4*7.1638E30/0.05*6.2E20/1E6*(470.0/600)*(470.0/600),
		600.0, 0, 1);


	auto heatmap = new TH2D("heatmap","heatmap; log(sin^2theta); log(Delta_m^2)",
				100,-3,0, 
				100,-2,2);

	for (int i = 1; i < heatmap->GetNbinsX()+1; i++)
	{
	    for (int j = 1; j < heatmap->GetNbinsY()+1; j++)
	    {
		float x = pow(10, heatmap->GetXaxis()->GetBinCenter(i));
		float y = pow(10, heatmap->GetYaxis()->GetBinCenter(j));

    	TH1D* h_sbnd2 = (TH1D*) h->Clone("h_sbnd2");
    	TH1D* h_microboone2 = (TH1D*) h->Clone("h_microboone2");
    	TH1D* h_icarus2 = (TH1D*) h->Clone("h_icarus2");

    	draw_histogram(h_sbnd2, cross_section, 
		    1E-38/1E4*1.6856E30/0.05*6.2E20/1E6*(470.0/110.0)*(470.0/110.0),
		    110.0, x, y);
    	draw_histogram(h_microboone2, cross_section, 
		    1E-38/1E4*1.3395E30/0.05*13.2E20/1E6,
		    470.0, x, y);
    	draw_histogram(h_icarus2, cross_section, 
	    	1E-38/1E4*7.1638E30/0.05*6.2E20/1E6*(470.0/600)*(470.0/600),
		    600.0, x, y);

    	double chi2_sbnd = chi2(h_sbnd, h_sbnd2);
	    double chi2_microboone = chi2(h_microboone, h_microboone2);
	    double chi2_icarus = chi2(h_icarus, h_icarus2);
	    double chi2_total = chi2_sbnd + chi2_microboone + chi2_icarus;
	    heatmap->SetBinContent(i,j,chi2_total);

	    std::cout << "chi2_sbnd: " << chi2_sbnd << std::endl;
	    std::cout << "chi2_microboone: " << chi2_microboone << std::endl;
	    std::cout << "chi2_icarus: " << chi2_icarus << std::endl;

	    std::cout << "x" << x << std::endl;
	    std::cout << "y" << y << std::endl;

	    delete h_sbnd2;
	    delete h_microboone2;
	    delete h_icarus2;
	    }	
	}
	    heatmap->Draw("colz");
	    gPad->SetLogz();
}
\end{lstlisting}

\end{document}